# TURBIDIMETRIC EVALUATION OF THE SOLUBILIZATION RATE: DISSOLUTION OF DODECANE NANODROPS IN 7.5 mM SODIUM DODECYLSULFATE SOLUTIONS AT SELECTED SODIUM CHLORIDE CONCENTRATIONS

José Daniel Rodríguez[1], Maurice Espinoza[3], Kareem Rahn-Chique[2], German Urbina-Villalba[2,*]

[1] Instituto Universitario Tecnológico "Dr. Federico Rivero Palacio", Carretera Panamericana, Km. 8, Caracas, Venezuela.

[2] Instituto Venezolano de Investigaciones Científicas (IVIC), Centro de Estudios Interdisciplinarios de la Física (CEIF), Carretera Panamericana Km. 11, Aptdo. 20632, Caracas, Venezuela. Email: german.urbina@gmail.com

[3] U.E. Los Hipocampitos, San Antonio de los Altos, Edo. Miranda, Venezuela.

**Abstract**  The rate of micelle solubilization ($S_R$) can be appraised following the decrease of the radius of a macroscopic drop of oil in contact with a surfactant solution [Todorov, 2002]. Alternatively, the time required for the dissolution of a liquid dispersion can be used for this purpose. Here, the decrease of the turbidity of a dodecane-in-water (d/w) nanoemulsion in 7.5 mM sodium dodecylsulfate (SDS) is studied at sodium chloride concentrations of 100, 300, 500, 700, 900, and 1000 mM NaCl. These salinities correspond to non-aggregating (< 300), aggregation-promoted (500) and surfactant precipitation regimes (> 700). It is found that $S_R \sim 2.3 \times 10^{-11}$, half the value observed in the absence of salt for a neat surfactant solution above its critical micelle concentration (7.0 < cmc < 8.7 mM SDS [Deodhar, 2020]).

**Keywords**  Emulsion, Nano, Dodecane, Solubilization, Rate, Measurement, CMC, Sodium Dodecylsulfate, Sodium Chloride.

## INTRODUCTION

The appearance of oil-in-water nanoemulsions has generated a renewed interest in the phenomenon of solubilization [Ariyaprakai, 2007; 2008] since micelles can completely disintegrate this type of dispersion in a few minutes [Rao, 2012]. Additional interest comes from the fact that common low energy methods for the synthesis of ionic nanoemulsions generally employ a high salt concentration in order to get the surfactant molecule out of the water phase, a key step of phase-inversion mechanisms. However, high ionic strengths reduce the critical micelle concentration (cmc) of the surfactant, leaving the resulting emulsion susceptible of solubilization. This is one of the reasons why a sudden dilution of the system is necessary at the end of the preparation process.

Solubilization basically depends on the surfactant concentration, the salinity of the aqueous solution, and the volume fraction of oil [Rahn-Chique, 2017]. These variables regulate the appearance of micelles and the amount of oil that can be dissolved.

Many-particle simulations *without* solubilization predict slower flocculation rates than experimentally found for 7.5 mM SDS nanoemulsions at [NaCl] ≤ 500 mM [Rahn-Chique, 2017; Urbina-Villalba, 2015; 2016]. Similar results are obtained for both dodecane-in-water and hexadecane-in-water (h/w) nanoemulsions. When a simple routine of solubilization is incorporated in the calculations [Rahn-Chique, 2017], it becomes evident that the outcome depends sensibly on: a) the solubilization rate: $S_R = dR/dt$ (where R is the average drop radius of the emulsion), and b) the maximum number of oil molecules that can be incorporated inside a micelle ($N_{oil}^{max}$). Former 25-particle simulations approximated these parameters by: the solubilization rate of the oils *in the absence of salt* (1.92 x $10^{-13}$ m/s for hexadecane, and 4.49 x $10^{-11}$ m/s for

José Daniel Rodríguez, Maurice Espinoza, Kareem Rahn-Chique, German Urbina-Villalba

dodecane [Ariyaprakai, 2008]), and the maximum solubilization capacity per micelle of **hexadecane** in water ( $N_{mic}^{oil} = N_{oil}^{max} = N_{h}^{max} \sim 2$ [Ariyaprakai, 2007; 2008]). However, it is known that $S_R$ markedly increases with the surfactant concentration [Todorov, 2002], and it is also the very likely that $N_{oil}^{max}$ changes with the ionic strength of the aqueous solution, since it sensibly increases with the surfactant concentration ($25 < N_d^{max} < 54$ for 35 mM < [SDS] < 139 mM [Ariyaprakai, 2008]), and the cmc of SDS decreases with the amount of sodium chloride [Urbina-Villalba, 2015].

In this regard, Mazer et al. [Mazer, 1976] studied the shape and size of SDS micelles at surfactant concentrations well above the cmc (17, 35 and 69 mM), NaCl concentrations between 150 and 600 mM, and temperatures of $10 \leq T \leq 85$ °C. With increasing temperature the SDS micelle asymptotically approaches a spherical shape with a hydrated radius of 25 Å nearly independent of NaCl and SDS concentrations. However, as the temperature is lowered in NaCl concentrations greater than 300 mM NaCl, micelles turn into prolate ellipsoids. For 69 mM SDS and 600 mM NaCl, their aggregation number ($N_{mic}^{surf}$) increases from 60 at T = 85 °C to 1600 at T = 18 °C. Alternatively, for the same surfactant concentration at T = 25 °C it is found that $N_{mic}^{surf}$ changes from 80 at 150 mM NaCl to 1000 in 600 mM NaCl. Ulterior examination of the data indicated that the lower temperatures examined surpass the solubility limit of the surfactant, and rather correspond to small crystallites [Hayashi, 1990]. Yet, Hayashi and Ikeda showed that SDS micelles formed at the cmc, aggregate into bigger structures above 450 mM NaCl (T = 25 °C). For 600 and 800 mM NaCl, the light scattering signal "exhibits an anomalous dissymmetry attributable to the formation of trace amounts of microgel particles". However, at T = 35 °C the scattering signal is normal again and indicates the presence of rodlike micelles with aggregation numbers as high as 1410 [Hayashi, 1990].

An order of magnitude estimation of $S_R$ can simply be obtained by direct observation of the phenomenon in test tubes. Figures 1 and 2 show the evolution of d/w nanoemulsions "stabilized" with 7.5 mM SDS at selected salt concentrations between 100 and 1000 mM NaCl. These emulsions have an average radius of R = 80 nm. According to previous studies, the flocculation rate is extremely slow for [NaCl] ≤ 300 mM, and therefore, solubilization prevails [Rahn-Chique, 2017]. Between 400 ≤ [NaCl] < 550 mM flocculation is the predominant phenomenon, and between 600 ≤ [NaCl] < 1000 mM surfactant precipitation takes place at room temperature (20 < T < 25) [Rahn-Chique, 2017; Díaz, 2018].

Inspection of Figure 1 indicates that a substantial clarification of 100 – 300 mM NaCl d/w emulsions occurs between t = 35 and t = 75 minutes. As a consequence $1.8 \times 10^{-11} < S_R = \Delta R / \Delta t = (80 \times 10^{-9} - 0) \text{ m} / (t - 0) \text{ s} < 3.8 \times 10^{-11}$ m/s. In previous simulations, a solubilization time ($t_s$) of 29 minutes was obtained assuming $S_R = 4.49 \times 10^{-11}$ m/s, $R_0 = 72.5$ nm and $N_d^{max} = 25$. The radius of the remaining particle at the end of the simulation was 4.8 nm, which is comparable to the average radius of a SDS micelle (2 nm [Cabane, 1985; Todorov, 2002; Reiss-Husson, 1959]). Interestingly, a sensible Ostwald ripening contribution was observed during the last 7 minutes of evolution (Fig. 19(b) in [Rahn-Chique, 2017]). This contribution diminishes the slope of dR/dt at constant solubization rate. It can be estimated that $S_R$ should be approximately equal to $3.7 \times 10^{-11}$ m/s in order to reproduce a value of $t_s$ = 35 min, since the velocity of solubilization is inversely proportional to $t_s$.

As confirmed by the calculations, a value of $N_d^{max} = 25$ is equivalent to assuming infinite solubilization capacity of the solution, since the amount of oil corresponding to $\phi = 3.2 \times 10^{-4}$ nanoemulsion can be completely absorbed by the available micelles. Conversely, if it is supposed that $N_d^{max} \sim N_h^{max} \sim 2$, the average radius of the emulsion diminishes during the first two minutes, but remains stable afterwards due to the saturation of the micellar solution (Fig. 19(a) in [Rahn-Chique, 2017]). This second scenario is not consistent with the experimental evidence.

Accurate Emulsion Stability Simulations (ESS) require reliable estimates of $S_R$ and $N_d^{max}$. Notice that a small cubic box (L = 70 $R_0$) containing 25 drops with a Gaussian distribution ($\phi = 3.06 \times 10^{-4}$) already contains $1.06 \times 10^8$ molecules of dodecane. Hence, a concentration of 7.5 mM SDS corresponds to approximately $9.8 \times 10^6$ micelles, excluding adsorbed surfactant molecules, and assuming a number of molecules per micelle ($N_{mic}^{surf}$) equal to 60 [Pisárčik, 2015; Todorov, 2002; Bales, 1998; La Mesa, 1990; Hayashi, 1990; Vass, 1988; Cabane, 1985]. Values of $N_{mic}^{surf}$ = 60, 50, 40 and 27 require $N_d^{max}$ = 11, 10, 8, and 5, respectively, in order to completely solubilize the 25 drops. Notice also that the progressive solubilization of the drops releases surfactants molecules adsorbed, which must form additional micelles if the monomer bulk concentration remains constant. In the present case, the number of surfactants adsorbed ($N_{surf}^{ads}$) is equal to $4.6 \times 10^6$. These molecules can form $7.7 \times 10^4$ micelles, and

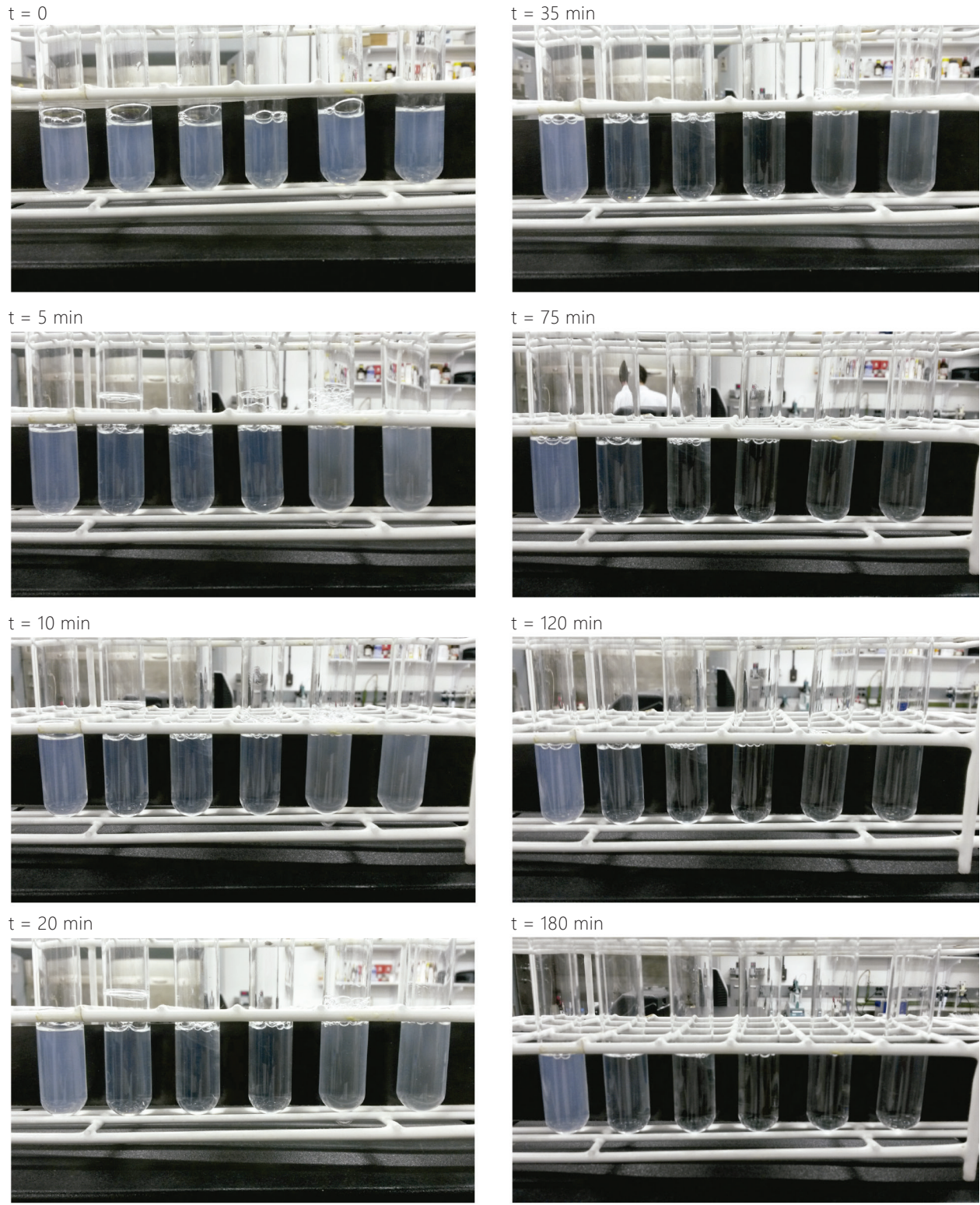

Fig. 1: Behavior of 7.5 mM SDS dodecane-in-water emulsions as a function of the salt concentration. From left to right test tubes correspond to: [NaCl] = 0, 100, 200, 300, 400, and 500 mM.

José Daniel Rodríguez, Maurice Espinoza, Kareem Rahn-Chique, German Urbina-Villalba

t = 0
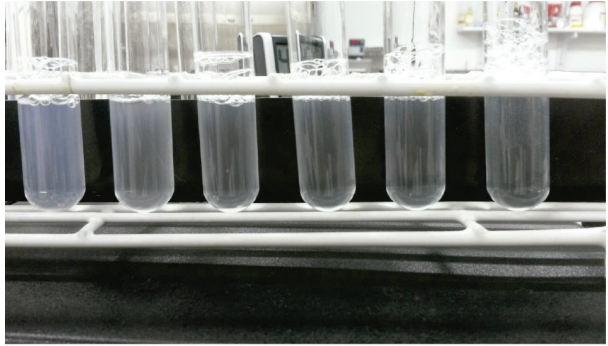

t = 30 min
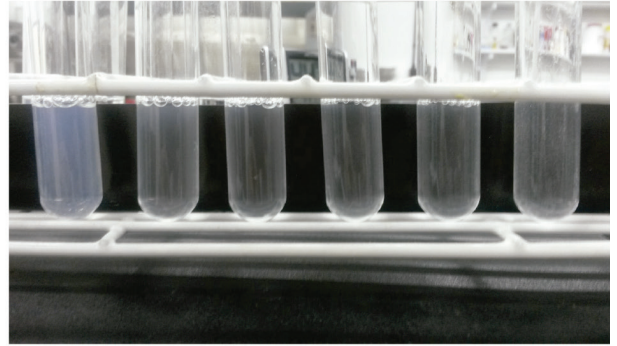

t = 5 min
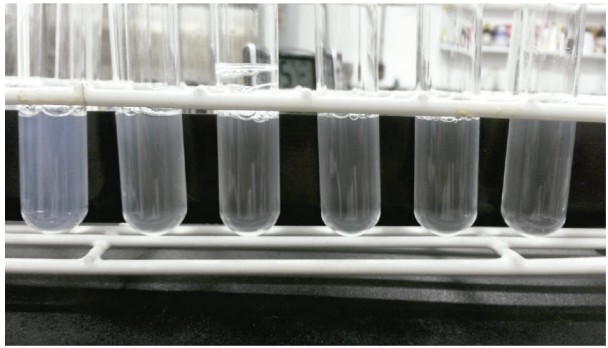

t = 90 min
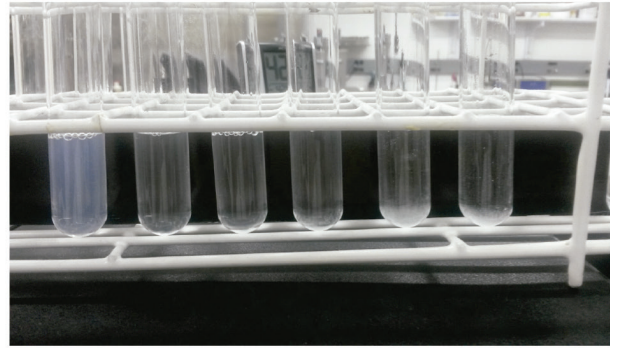

t = 15 min
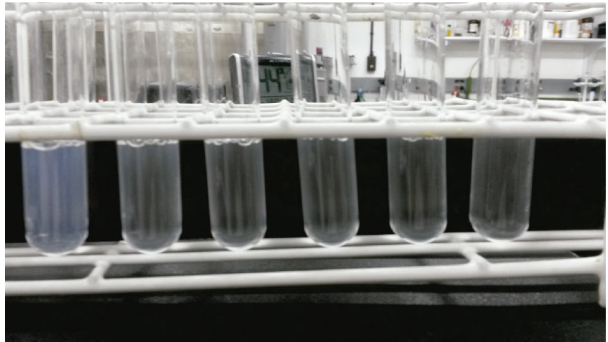

t = 120 min
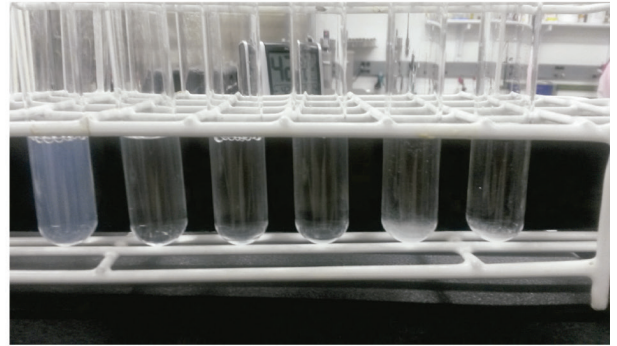

t = 20 min
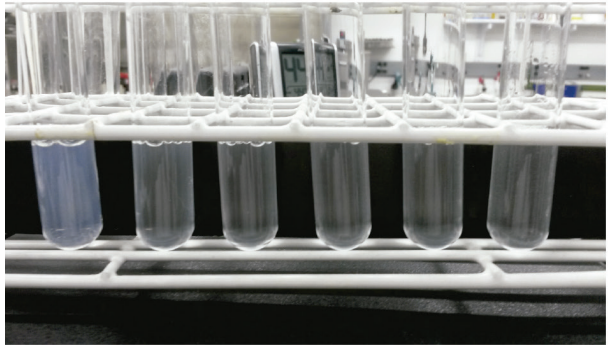

t = 240 min
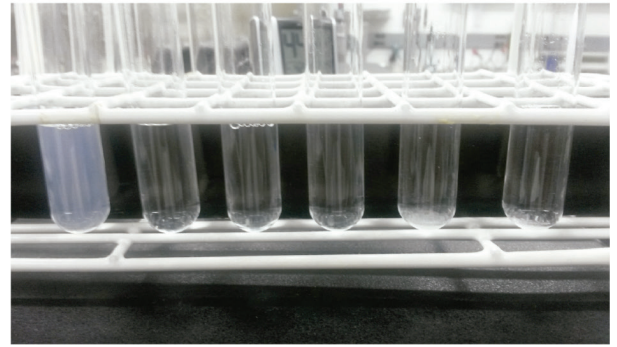

Fig. 2: Behavior of 7.5 mM SDS dodecane-in-water emulsions as a function of the salt concentration. From left to right test tubes correspond to: [NaCl] = 0, 600, 700, 800, 900, and 1000 mM.

solubilize an additional amount of 8.5 x 10$^5$ molecules of dodecane for $N_d^{max}$ = 25.

It is clear from these numbers that it is not possible to use molecular simulations to reproduce the processes of micellar solubilization (and/or Ostwald ripening) along with the aggregation of drops. Not even for drops of nanometric size. A similar technical difficulty is confronted with surfactant adsorption. This process is also time dependent, and moderates the repulsive interaction between the drops. Furthermore, the time step of the simulations has to be short enough to sample the potential of interaction appropriately. A typical repulsive barrier occurs at a distance of a few nanometers from the particle's surface. This limitation typically restricts Δt to values of the order of $10^{-8}$ s.

The test tubes of Figures 1 and 2 also illustrate the difficulties of evaluating $S_R$ when other destabilization phenomena concurrently occur. Clarification also happens if the suspended particles migrate out of the bulk of the solution. This can occur if the drops flocculate and cream out, or if the surfactant crystallizes and then precipitates. During a transient period, the decrease of the turbidity caused by solubilization opposes the increase promoted by either aggregation or crystal nucleation. However, after a transient period these additional mechanisms also contribute to the clarification of the sample due to creaming or precipitation. In these cases, the use of the turbidity for the evaluation of the solubilization rate might be difficult.

In a previous work [Rahn-Chique, 2017] the maximum volume fraction of oil that can be dissolved by a 7.5 mM SDS solution ($\phi = \phi_c$) was determined for several concentrations of sodium chloride. This volume fraction is implicitly a function of time, since it depends on $S_R$: $\phi_c = \phi_c(t)$

The maximum number of oil molecules that can be dissolved by a micelle ($N_{oil}^{max}$) can be deduced using $\phi_c(t)$ or $\phi_c(t=\infty)$. By definition, the volume fraction of oil in an oil-in-water emulsion is equal to:

$$\phi = \frac{V_o}{V_o + V_w} = \frac{V_o}{V_T} \quad (1)$$

Where $V_o$ and $V_w$ are the volumes of oil and water. When, the critical volume fraction is achieved ($\phi = \phi_c$), all molecules of oil ($N_{oil}^{molec}$) are distributed amongst the existing number of micelles ($N_{mic}$):

$$N_{oil}^{molec} = N_{mic}^{oil} N_{mic} \quad (2)$$

Where $N_{mic}^{oil}$ is the average number of oil molecules dissolved in a micelle. At $\phi = \phi_c$, this value corresponds to the maximum number of molecules that can be solubilized by a micelle $N_{oil}^{max}$ [Rahn-Chique, 2017]. Therefore:

$$V_o = \phi_c V_T = N_{oil}^{molec}\left[\frac{MW_o}{N_A d_o}\right] = N_{oil}^{max} N_{mic}\left[\frac{MW_o}{N_A d_o}\right] \quad (3)$$

Where $MW_o$ (g/mol) and $d_o$ (g/ml) are the molecular weight and the density of the oil, and $N_A$ is Avogadro's number.

The maximum number of micelles ($N_{mic}$) is equal to the total number of surfactant molecules ($N_{surf}$) divided by the number of surfactants per micelle ($N_{mic}^{surf}$):

$$N_{mic} = \frac{N_{surf}}{N_{mic}^{surf}} = \frac{C_s V_T}{N_{mic}^{surf}} \quad (4)$$

Where $C_s$ is the total surfactant concentration. Equation (4) applies to the surfactant solution only, otherwise the concentration of surfactant molecules adsorbed to the interface of the drops ($N_{surf}^{ads}$), or dissolved as monomers in the bulk of the solution ($N_{surf}^{mono}$) should be subtracted from the total surfactant population:

$$N_{mic} = \frac{N_{surf} - N_{surf}^{ads} - N_{surf}^{mono}}{N_{mic}^{surf}} \approx \frac{C_s V_T}{N_{mic}^{surf}} \quad (5)$$

From Eq. (3) and (4):

$$\phi_c \approx \frac{1}{V_T} N_{oil}^{max} N_{mic}\left[\frac{MW_o}{N_A d_o}\right] = \frac{1}{V_T}\left[\frac{N_{oil}^{max} N_{surf}}{N_{mic}^{surf}}\right]\left[\frac{MW_o}{N_A d_o}\right]$$

$$= \frac{1}{V_T}\left[\frac{N_{oil}^{max}}{N_{mic}^{surf}}\right][V_T C_s N_A][10^{-3} \text{ l/ml}]\left[\frac{MW_o}{N_A d_o}\right] \quad (6)$$

In Eq. (6) it is supposed that the surfactant concentration is in moles/l, and therefore, factor $N_A[10^{-3}$ l/ml$]$ is required to transform it into molecules / ml. Finally:

$$\phi_c = \left[\frac{N_{oil}^{max}}{N_{mic}^{surf}}\right][C_s]\left[\frac{MW_o}{d_o}\right][10^{-3} \text{ l/ml}] \quad (7)$$

Or:

$$N_{oil}^{max} = \frac{\phi_c d_o N_{mic}^{surf}}{C_s MW_o}[10^{+3} \text{ ml/l}] \quad (8)$$

José Daniel Rodríguez, Maurice Espinoza, Kareem Rahn-Chique, German Urbina-Villalba

Equation (8) allows calculating the maximum number of oil molecules dissolved in a micelle if the critical volume fraction of oil is measured and $N_{mic}^{surf}$ is known. Using the values of $\phi_c$ reported in [Rahn-Chique, 2017] for T = 25 and T = 20 °C, the physical properties of dodecane (MW$_o$ = 170.34, d$_o$ = 0.749 g/ml, C$_s$ = 7.5 mM, and assuming $N_{mic}^{surf} = N_{mic}^{SDS} \approx 60$, the entries of Table 1 are obtained. These numbers are very interesting: For t ≤ 10 minutes the number of oil molecules per micelle is small, between 1 and 6. However, at very long times: $16 \leq N_d^{max} \leq 66$ for $100 \leq [NaCl] \leq 500$ mM. These figures decrease considerably for salinities that promote surfactant precipitation: [NaCl] > 500 mM. The critical volume fraction does not change monotonously with the amount of salt, but for T = 25 °C: $N_d^{max}$ decreases with the ionic strength of the aqueous phase.

Previous simulations only predict complete solubilization of a d/w emulsion with $\phi = 3 \times 10^{-4}$ if $N_d^{max} > 11$. This inconsistency is probably due to the fact that this volume fraction is larger than $\phi_c$ (see Tables 2 and 3 in Ref. [Rahn-Chique, 2017]).

It is remarkable that the quotient $N_{mic}^{oil}/N_{mic}^{surf}$ (fraction of oil molecules per surfactant in a micelle) is proportional to $\phi_c$ (Eq. (7)) since it is well established that as the surfactant concentration increases, the number of micelles and their size increase.

If experimental solubilization rates are lower than expected in the absence of salt, it is likely that the transfer of oil occurs via molecular solubilization of oil in the water phase, with subsequent adsorption of the oil molecule by a micelle in the bulk of the solution. Since the solubility of the oil decreases with the ionic strength, so will the rate of solubilization [Rahn-Chique, 2017]. Instead, if micelles attach to the surface of the drops and withdraw the oil molecules directly, the electrostatic repulsion between micelles and drops should decrease with the salinity, and a subtle increase of the solubilization rate with the ionic strength is predictable.

The present report attempts to use turbidity measurements in order to appraise the value of $S_R$ in the presence of significant amounts of salt.

**EXPERIMENTAL PROCEDURE**

Dodecane (Merck, 98%) was purified using an alumina column. Sodium dodecyl sulfate (SDS), sodium chloride (Merck, 99.5%) and isopentanol (Scharlau Chemie, 99%) were used as received. Distilled water was deionized using a Millipore Simplicity apparatus.

Table 1: Maximum number of dodecane molecules per micelle at C$_s$ = 7.5 mM SDS.

| [NaCl] mol/l | T (°C) | $\phi_c$ | Time (t) of observation | $N_{oil}^{max}$ |
|---|---|---|---|---|
| 100 | 25 | 3.27 x 10$^{-5}$ | 10 min | 1 |
| 300 | 25 | 1.13 x 10$^{-4}$ | 10 min | 4 |
| 500 | 25 | 2.71 x 10$^{-5}$ | 10 min | 1 |
| 700 | 25 | 1.74 x 10$^{-4}$ | 10 min | 6 |
| 900 | 25 | 9.80 x 10$^{-5}$ | 10 min | 3 |
| 100 | 25 | 9.75 x 10$^{-4}$ | 7 days | 34 |
| 300 | 25 | 1.36 x 10$^{-3}$ | 7 days | 48 |
| 500 | 25 | 4.60 x 10$^{-4}$ | 7 days | 16 |
| 700 | 25 | 2.62 x 10$^{-4}$ | 7 days | 9 |
| 900 | 25 | 1.57 x 10$^{-4}$ | 7 days | 6 |
| 100 | 20 | 3.04 x 10$^{-5}$ | 10 min | 1 |
| 700 | 20 | 1.48 x 10$^{-4}$ | 10 min | 5 |
| 100 | 20 | 1.88 x 10$^{-3}$ | 7 days | 66 |
| 300 | 20 | 6.62 x 10$^{-4}$ | 7 days | 23 |

Nanoemulsions were prepared using a phase inversion composition method [Rahn-Chique, 2012a]. A mixture of liquid crystal solution and oil ($\phi$ = 0.84, [SDS]= 10 wt%, [NaCl]= 8% wt%, and [isopentanol]= 6.5 wt%) previously pre-equilibrated, was suddenly diluted until $\phi$ = 0.44, to obtain 120-150 nm drops of oil. This mother nanoemulsion was further diluted successively with pure water until $\Phi$ = 0.02, and with an appropriate surfactant solution until $\phi = 3.2 \times 10^{-4}$.

The initial radius of the mother emulsions was appraised employing a Beckman-Coulter LS-230 particle analyzer. The variation of the average radius as a function of time was observed for a selected number of systems using a Goniometer from Brookhaven Instruments.

A Quickscan™ turbimeter (Beckman-Coulter-Formulaction) was used to measure the Transmittance (Tr) of the emulsions as a function of time. For every experiment the Tr of pure water was measured first. The automatic evaluation of each system started by measuring a sample of freshly prepared emulsion without salt. In between measurements 1 and 2, the former tube was taken out of the apparatus, and a third tube prepared mixing the appropriate volumes of emulsion and salt solution. This new measuring cell was set in place in the Quickscan, prior to the second automatic measurement. From this time on, the transmittance was measured periodically in the usual manner until maximum clarification of the sample was achieved (2 - 24 hours depending on the amount of salt).

This turbimeter uses an 850-nm light source and two synchronous detectors to evaluate the light transmitted at θ = 0°, and reflected at θ = 135° along the height of a measuring cell (every 40 μm). The intensity of light depends on the size of the suspended particles and its volume fraction. For example, for a fixed volume fraction of 1% of latex beads on water, the logarithm of the total reflectance (Re) increases with a slope of 3/2 with respect to the diameter of the particles (d) when d < 280 nm, but decreases with slope of 1/2 above this size (d > 280 nm). Since Tr = 1 – Re, the reverse variation is expected when the transmittance is appraised. Hence, Tr should increase as the size of the drops decrease.

Most samples were initially Gaussian or log-normal distributions of emulsion drops with an average radius around 60 nm and a typical standard deviation of 700 nm (!). These systems develop either micelles or crystallites depending on the salt concentration and temperature. Crystals can grow up to macroscopic sizes and precipitate, but swollen micelles are too small to show a sensible gravitational effect. The volume fraction ($\phi \sim 3.2 \times 10^{-4}$) is the one previously employed in customary flocculation experiments. The initial samples look transparent but hazy (Figs. 1 and 2).

Figure 3 illustrate the long time behavior of the surfactant solutions at different salt concentrations. The presence of either micelles or crystals at equilibrium depends on the Krafft temperature, which varies between 20-25 °C for aqueous solutions of SDS. These temperatures are typical room temperatures of our lab.

In the absence of drops, the amount of surfactant lost by absorption is negligible, and crystal formation is observed even at 400 mM NaCl after 24 hours. In the presence of a dilute emulsion ($\phi = 3.2 \times 10^{-4}$) a substantial amount of surfactant is lost by adsorption. This process sensibly diminishes the quantity of SDS available for crystal formation. Thus, in the presence of emulsions (Figs. 1 and 2), surfactant precipitation is only observed for [NaCl] ≥ 900 mM after 90 minutes, and for [NaCl] ≥ 700 mM when t→∞. It must be kept in mind that the cmc diminishes from 1.62 to 0.56 mM SDS as the amount of sodium chloride changes from 100 to 1000 mM [Urbina-Villalba, 2016]. Hence, the amount of surfactant in solution available for micelles or crystals increases with the salt concentration. In all cases, the amount of surfactant adsorbed to the drops is much lower, though it increases with the salt concentration due to the salting out of SDS.

Conventional destabilization processes (flocculation and coalescence) lead to an absorbance increase (lower Tr). Instead, sedimentation and solubilization promote a clarification of the sample. Oswald ripening can work either way, diminishing the average size of the drops during the exchange of molecules, and increasing it when drops are eliminated along the process. Thus, its influence on light scattering is more difficult to ascertain. It changes the shape of the drop size distribution sensibly, and this influences the scattering process.

Sedimentation and creaming are easily identified with a turbimeter of variable height: Quickscan or Turbiscan Classic (Formulaction), due to the opposite variation of the scattering at the extremes of the sample tube. If precipitation occurs, the bottom of the tube gets increasingly obscure, and its top clarified. Hence, the sample develops peaks of minimum and maximum Tr at the edges of the container. The opposite occurs when creaming takes place. For micelles and for drops of nanometer size, the influence of gravity is expected to be negligible, unless a substantial aggregation occurs.

As shown in Figs. 1 and 2, the process of solubilization increases Tr -along the complete length of the tube- until clarification. In this case a plot of Tr vs. h (where h is the height of the measuring cell) should show parallel lines which increase as a function of time from a minimum Tr (corresponding to the cloudy emulsion) until complete clarification. Consequently, at a constant height, a smooth increase of Tr vs. t is expected, unless crystallization occurs. However, it was found that Tr generally shows an initial decrease and a minimum value, followed a sharp increase and a final asymptotic trend with fluctuations around a maximum value (see Figure 4). This behavior occurred despite the specific value of h selected for the analysis. In this work three zones of analysis were chosen along the length of the sample tube to calculate average values of Tr: Zone 1: 8.5 - 13.5 mm, Zone 2: 25-26 mm, and Zone 3: 35.5 – 40.5 mm. All plots shown in this paper correspond to the data collected for Zone 1 (bottom of the tube). These zones should not be confounded with the regions of variation of the average values of Tr as a function of *time* (see below).

In order to study the behavior of the systems, a minimum of three linear slopes were drawn on each plot of Tr vs. t. They correspond to the two sides of the minimum, and the final trend, respectively (see Fig. 4). The solubilization time $t_s$ was defined as the time corresponding to the

José Daniel Rodríguez, Maurice Espinoza, Kareem Rahn-Chique, German Urbina-Villalba

t = 0

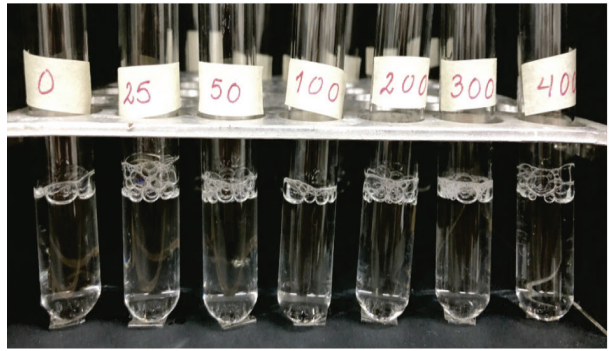 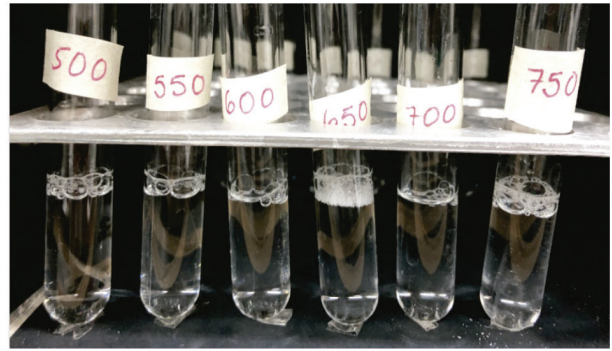

t = 10 min

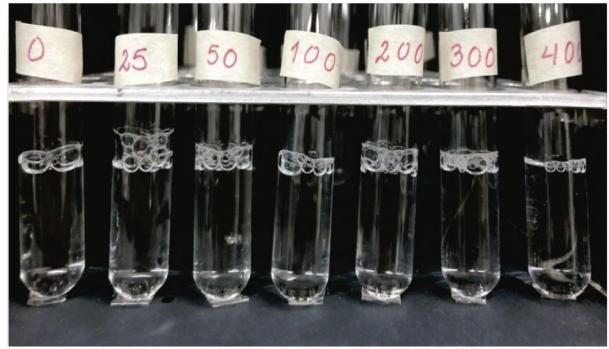 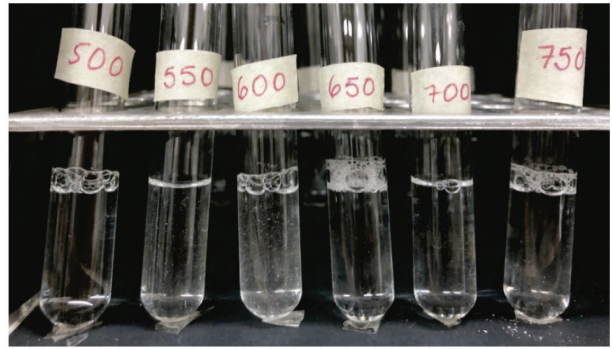

t = 30 min

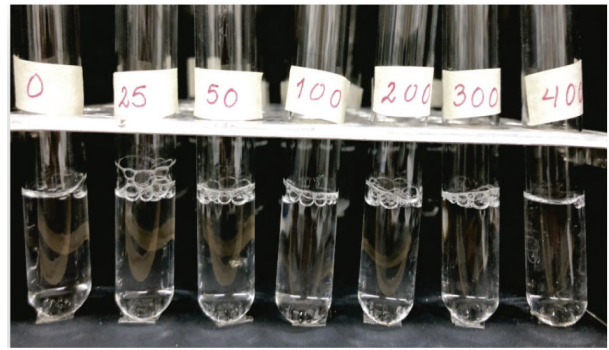 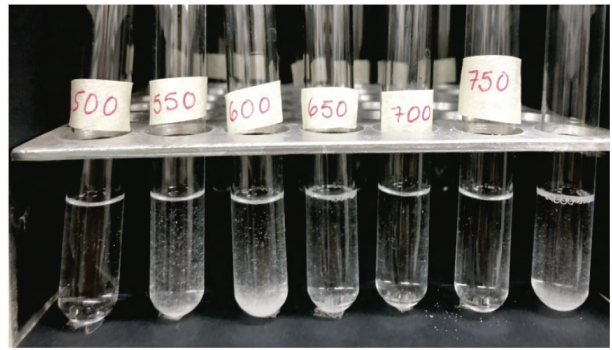

t = 60 min

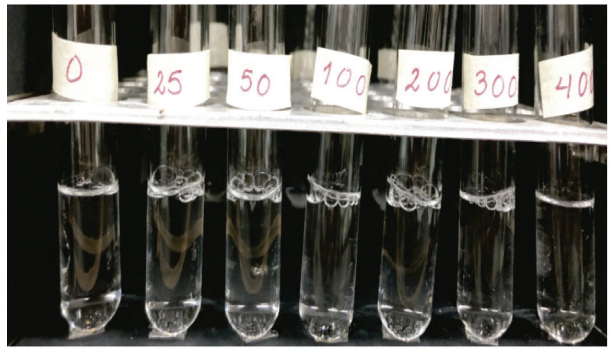 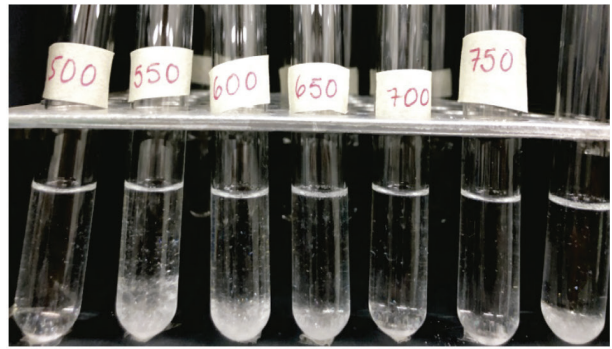

Fig. 3(a): Phase behavior of 7.5 mM SDS solutions as a function of the salt concentration. The labels of the tubes correspond to [NaCl] in mM.

t = 1440 min (24 h)   0.5 mM SDS

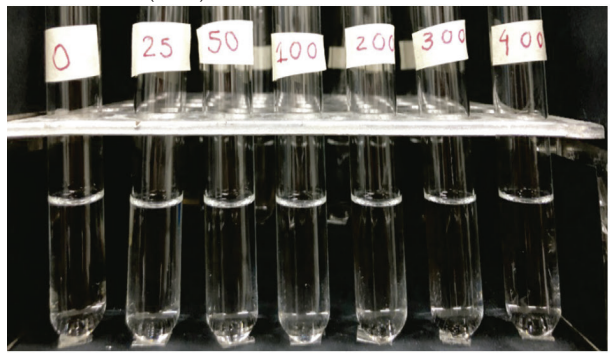
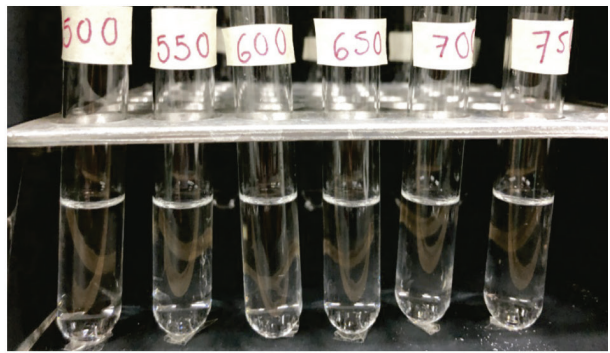

t = 1440 min (24 h)   7.5 mM SDS

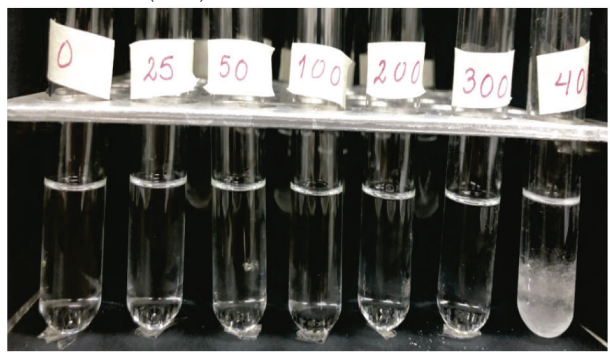
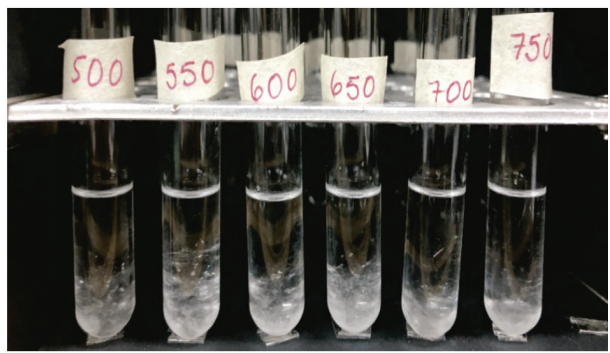

Fig. 3(b): State of 0.5 and 7.5 mM SDS solutions 24 hours after preparation.
The labels of the tubes correspond to [NaCl] in mM.

intersection between the last two lines (the one defining the initial sharp increase and the one corresponding to the asymptotic slope). It is assumed that the initial increase mostly reflects the effect of solubilization, while the final one is severely affected by ripening. According to previous simulations [Rahn-Chique, 2017], Oswald ripening contributes significantly at the end of the solubilization process promoting a sharp change of slope of dR/dt (see Fig. 19 in [Rahn-Chique, 2017]).

The time of required for the end of the clarification process: $t_c$, was estimated using the last slope of Tr vs. t, and its actual increase. It is defined as the *minimum* time required for Tr to reach the constant value of the last linear regression (b for Tr = a t + b). Since Tr markedly fluctuates at long times, the intercept of the line with the vertical axis defines a kind of minimum value for the beginning of systematic fluctuations which are typical of the Ostwald ripening process [Urbina-Villalba, 2009].

A more conventional clarification time was defined as the minimum time required for Tr to vary less than a certain

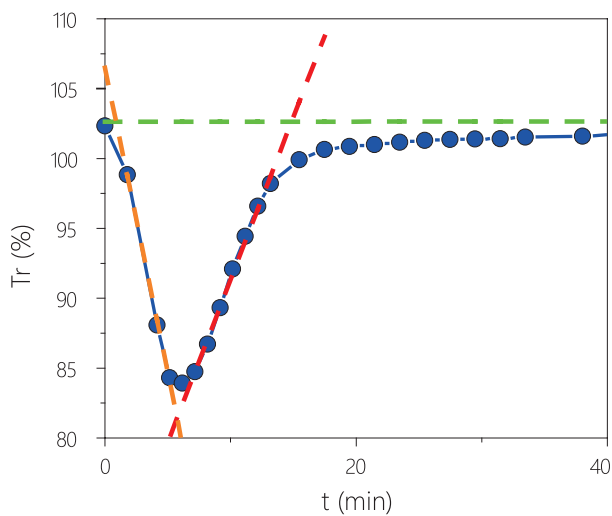

Fig. 4: Typical Transmittance vs. time curve. Time $t_s$ results from the intersection of two straight lines: the one corresponding to the fitting of the long-time values of Tr, and the one representing the steep rise immediately after the minimum (which presumably corresponds to micelle solubilization). Time $t_c$ is the shortest time it takes for Tr to reach or surpass the long-time line. It is meant to appraise the end of the clarification process.

José Daniel Rodríguez, Maurice Espinoza, Kareem Rahn-Chique, German Urbina-Villalba

percentage until the end of the measurement. This can be evaluated from direct observation of the Turbimeter data. Clarification times of 0.5% ($t_{0.5}$) and 1.0% ($t_{1.0}$) were evaluated. Similar criteria of variation can be defined between two successive measurements. The advantages and drawbacks of each definition will be evident from the results.

The behavior of the transmittance was studied in three systems: dodecane/water (d/w) emulsions, (dodecane + 10% squalene)/water emulsions (dsq/w), and surfactant solutions (sw). On the one hand, addition of squalene to dodecane stops the process of ripening [Cruz-Barrios, 2014]. On the other hand, surfactant solutions were expected to serve as blanks to identify the contribution of the surfactant phase behavior to the transmittance of the emulsified systems. It is well known that SDS solutions only scatter light above the cmc.

In order to inquire into the nature of the processes which contribute to each section of the curve of Tr vs. t (Fig. 4) two approximate relations between Tr and the average radius of the emulsions (R) were considered:

$$Tr = \frac{k_1}{R^m} \quad (9)$$

$$Tr = 100 - k_2 R^m \quad (10)$$

These relations allow deducing the rate of change of the average radius. In the first case:

$$\frac{dTr}{dt} = \frac{-m k_1}{R^{m+1}} \frac{dR}{dt} \rightarrow \frac{dR}{dt} = \frac{-R^{m+1}}{m k_1} \frac{dTr}{dt} \quad (11)$$

In the second case:

$$\frac{dTr}{dt} = -m k_2 R^{m-1} \frac{dR}{dt} \rightarrow \frac{dR}{dt} = \frac{-1}{m k_2 R^{m-1}} \frac{dTr}{dt} \quad (12)$$

These expressions allow calculating the rate of change of the average radius. They only require the value of R at the start of the process (R = R($t=t_{0p}$)) and the average slope of Tr vs. t in the section (period) of interest (Fig. 4). They take into account that Tr increases as the radius of the drops decreases (Eq. 9), and that both experimental studies and simulations indicate that for sufficiently small drops the total reflectance (backscattering signal) is proportional to some power of the radius (For latex particles $Tr \propto R^{3/2}$ or $R^{1/2}$ according to [Beckman-Coulter, 2000]).

Additionally, dR/dt allows calculating the change of the cube average radius:

$$\frac{dR^3}{dt} = 3R^2 \frac{dR}{dt} \quad (13)$$

This rate of change can be compared with the value predicted for Oswald ripening by the theory of LSW [Lifshitz, 1961; Wagner, 1961]:

$$V_{OR} = \frac{dR^3}{dt} = \frac{8}{9} \frac{\gamma_{cmc} V_m D_m C_\infty}{R_g T} \quad (14)$$

Where $D_m$ = 5.40 x $10^{-10}$ $m^2$/s, $C_\infty$ = 5.31 x $10^{-9}$ ml/ml, $V_m$ = 2.27 x $10^{-4}$ $m^3$/mol, $R_g$ = 8.31 J/mol K, T = 298.13 K. In the absence of salt, and for [SDS] = 8 mM, $\gamma_{cmc}$ = 9.16 mN/m [Cruz-Barrios, 2014]. In this case $V_{OR}$ = 2.1 x $10^{-27}$ $m^3$/s. More accurate estimations can be obtained from the equation of Aveyard [Aveyard, 1987; Urbina-Villalba, 2013]:

$$\gamma_{cmc} = -0.77969 \ln([NaCl] + [SDS]) + 3.1458 \quad (15)$$

Using Eq. (15) with [SDS] = 7.5 mM, values of: 7.3 x $10^{-28}$ $m^3$/s ≤ $V_{OR}$ ≤ 1.1 x $10^{-28}$ $m^3$/s are obtained for 100 mM ≤ [NaCl] ≤ 1 M (4.9 mN/m ≤ $\gamma_{cmc}$ ≤ 3.1 mN/m).

**RESULTS AND DISCUSSION**

The present work can be regarded as preliminary since the turbimeter employed does not regulate the sample temperature, and gets appreciably warm with time. Since the Krafft point of SDS lies between 20 and 25 degrees, we kept the apparatus in a small air conditioned room maintained at T = 17-18 °C during the measurements. This provision increased considerably the reproducibility of the results, but the variation of the actual temperature of the samples is still unknown. Duplicate evaluations were made for each salt concentration, and additional measurements were endeavored when a lack of precision was evident.

Figures 5 and 6 show the behavior of two sets of surfactant solutions for 300, 500 and 700 mM NaCl. For 100 ≤ [NaCl] ≤ 1000 mM, the critical micelle concentration lowers to 1.62 ≤ cmc ≤ 0.56 mM. Assuming an average area per surfactant close to 50 Å²/molec, the adsorbed surfactant

Turbidimetric evaluation of the solubilization rate: Dissolution of dodecane nanodrops in 7.5 mM sodium dodecylsulfate solutions at selected sodium chloride concentrations

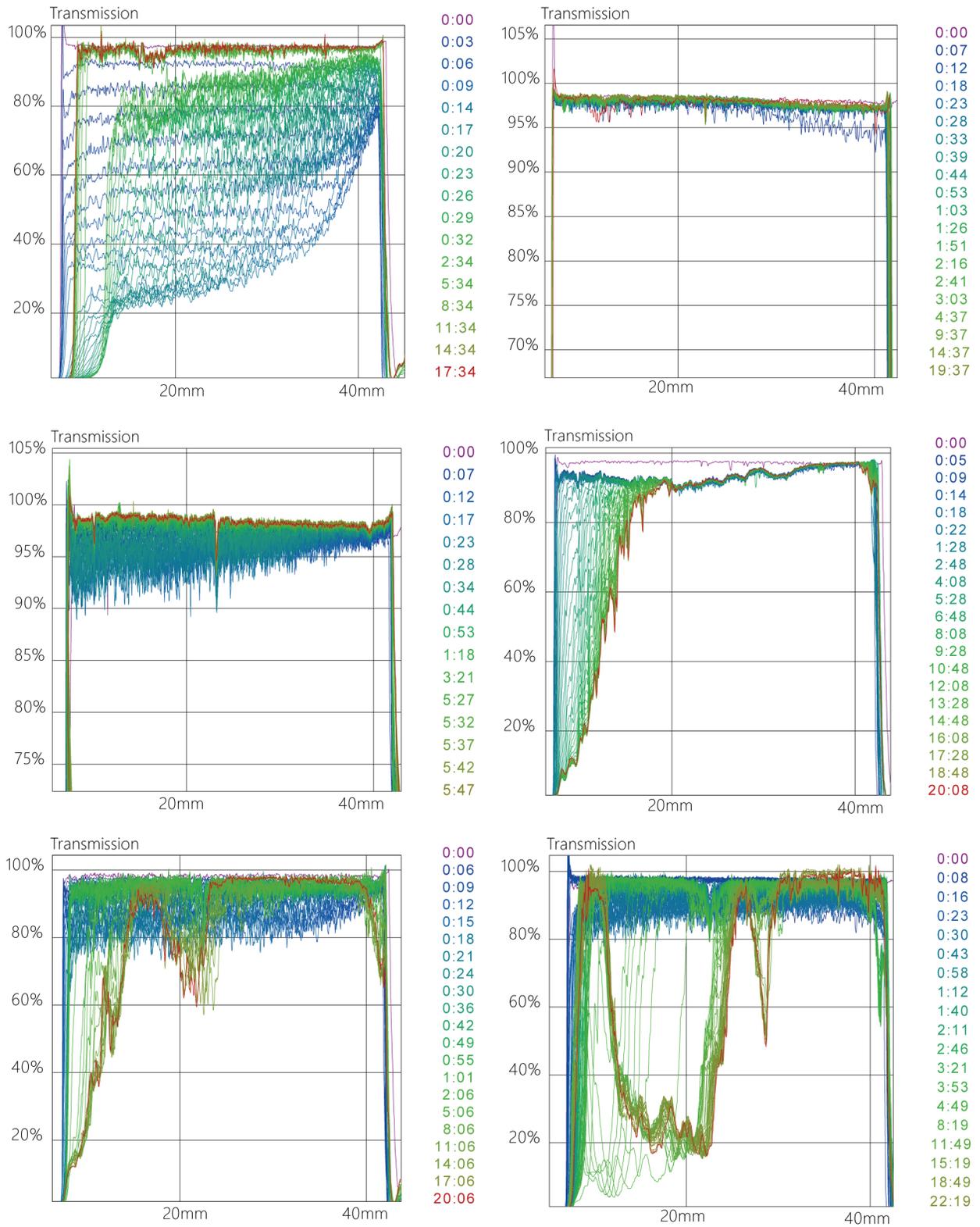

Fig. 5: Evolution of the curves of Tr vs h for **two sets** of 7.5 mM **SDS solutions** with 300 (first row), 500 (second row), and 700 mM NaCl. Temperature and mixing technique affect the formation of the micelles and the crystallization process.

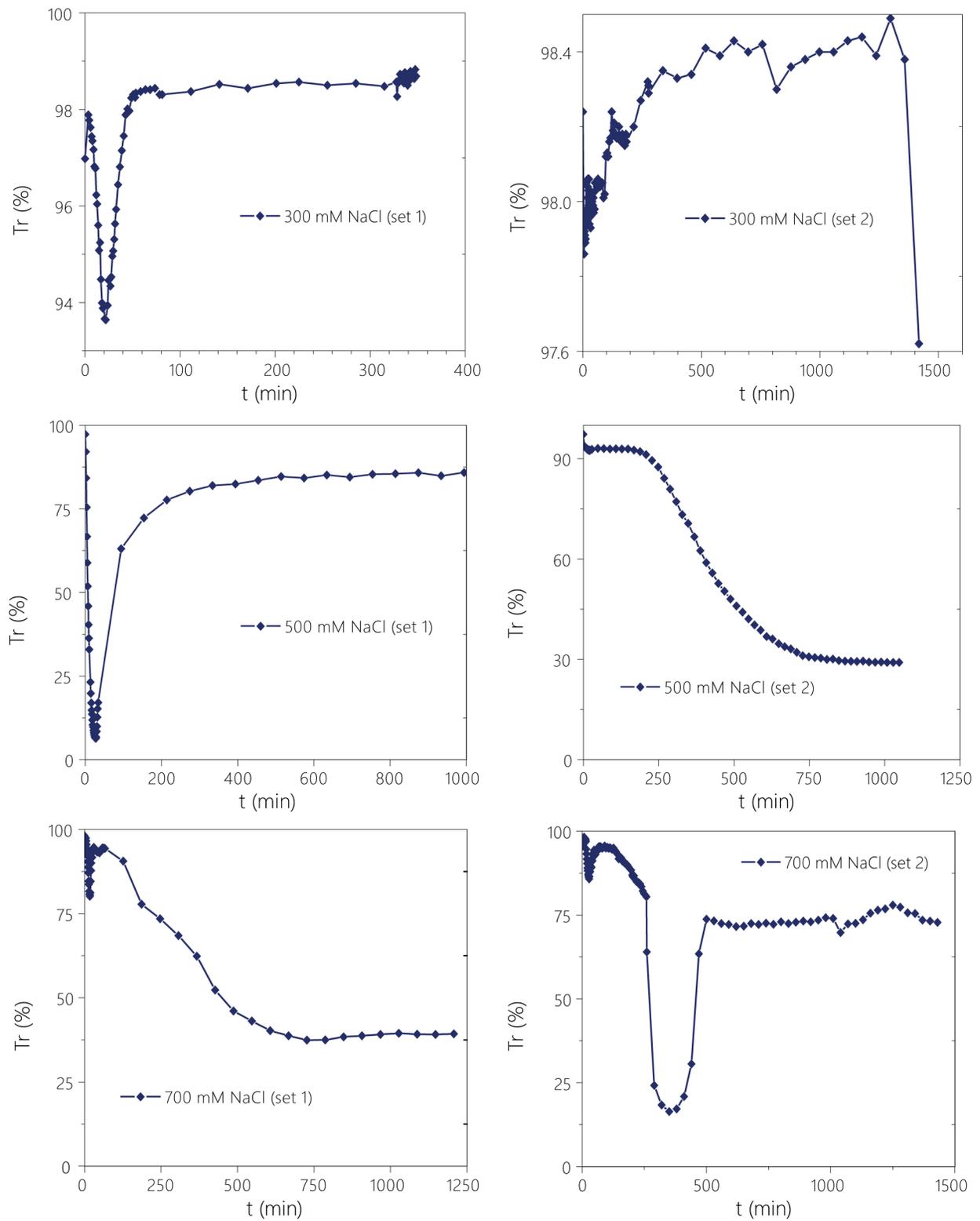

Fig. 6: Plots of Tr vs. t corresponding to the data of Tr vs. h shown in Figure 5.

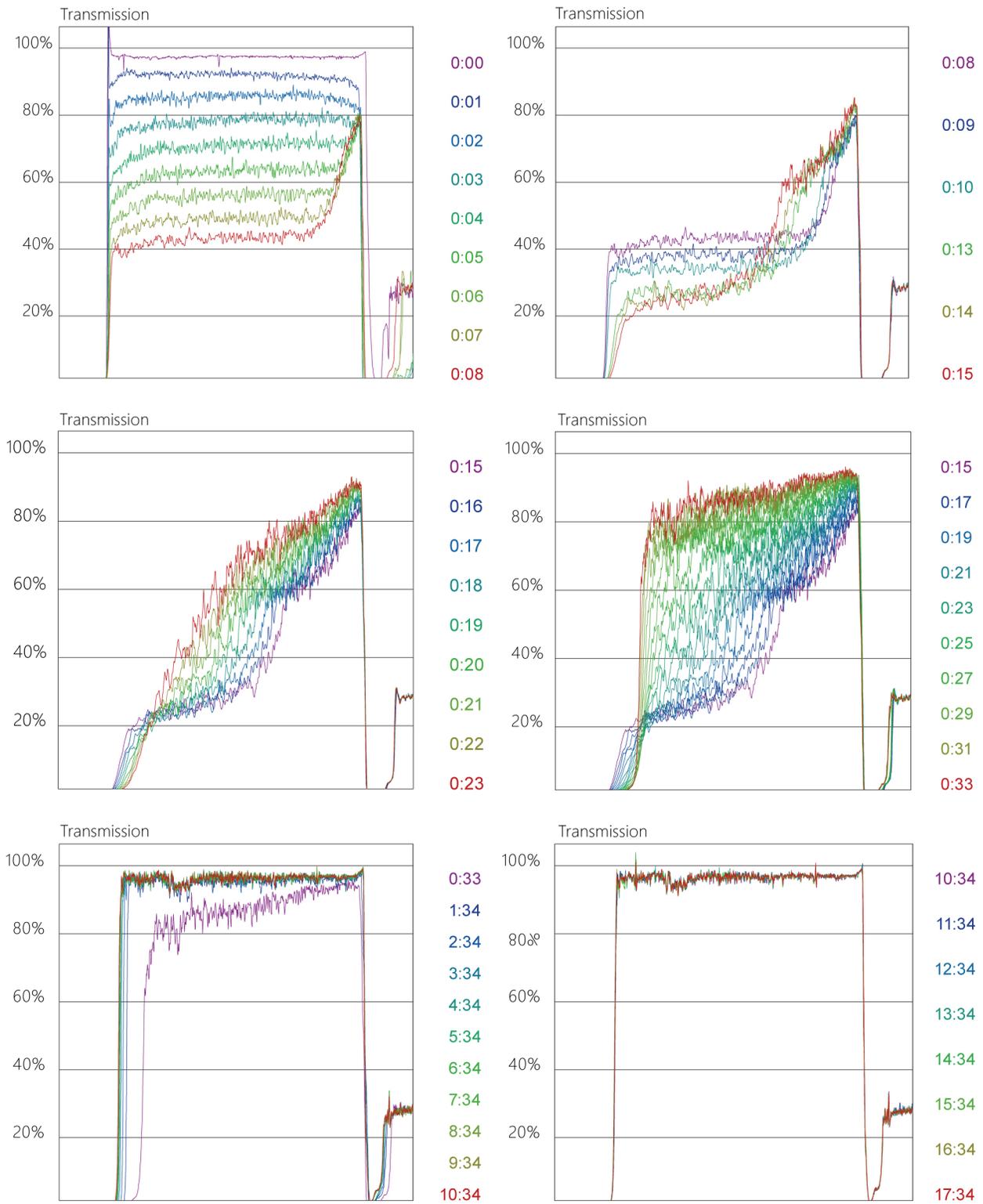

Fig. 7: Details of the evolution of Tr vs. h for a solution of 7.5 mM SDS with 500 mM NaCl. In each inset the violet and red curves correspond to the shortest and the longest times measured.

José Daniel Rodríguez, Maurice Espinoza, Kareem Rahn-Chique, German Urbina-Villalba

concentration becomes 0.04 mM for drops of R = 72.5 nm (at $\phi = 3.2 \times 10^{-4}$). This leaves a very large amount of surfactant molecules in solution: $5.84 \leq [SDS] \leq 6.90$ mM for the salt concentrations tested. These molecules can either form micelles or crystals depending on the actual temperature of the sample tube. It is evident that the transmission "spectra" of the turbimeter varies considerably depending on weather crystals precipitate or not. In the former case dips of Tr vs. h as profound as 20% develop in the regions of precipitation, and the plots of Tr vs. t behave erratically.

Although our instrument is old, its software is very versatile. It allows averaging of the Tr data over specific regions of the sample tube. We studied the bottom (8.5 – 13.5 mm), the middle (25-26 mm) and the top (35.5 – 40.5 mm) of the tube. For simplicity all curves of Tr vs t presented in this paper correspond to the bottom of the tube (Fig. 6).

Figure 7 shows the evolution of the curves of Tr vs. h for 500 mM NaCl. In each inset violet and red curves correspond to the shortest and longest times exhibited. After a few minutes the bottom of the tube gets obscure, and the top clarifies. As time evolves, the clarification zone widens spanning the whole length of the tube. In this particular case, the precipitate is amorphous (powder-like) and forms a thin layer at the base of the tube. In other cases (700 mM) wide regions of a vitreous-like phase are formed. The difference between these scenarios is evident by direct inspection of the sample tubes (see the cover illustration of RCEIF vol. 9).

When dodecane drops are present (d/w systems), a substantial part of the surfactant population is absorbed to the interface of the drops. This alters the equilibrium of the bulk solution. Figure 3 shows that in the absence of drops (sw systems) crystal formation is observed after 1 hour at 550 mM NaCl, and at 400 mM after 24 hours. In the case of d/w emulsions crystallization is only observed after 2 hours when $[NaCl] \geq 700$ mM, and above $[NaCl] \geq 500$ mM only after 24 hours.

As illustrated in Figure 8, short-time aggregation of d/w emulsions is only observed for [NaCl] > 300 mM. Thus, it is expected that the most reliable values of $S_R$ are obtained for 100 and 300 mM NaCl systems. A salinity of 500 mM causes aggregation of the drops. This is evidenced by the monotonous increase of the average radius. The appearance of crystals (inset of Fig. 8) is characterized by substantially higher values and the appearance of peaks, which resemble the characteristic coordination spheres of a radial distribution function (g(r)).

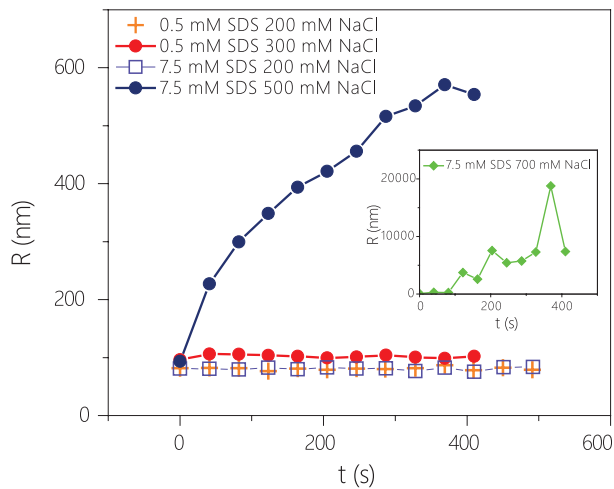

Fig. 8: Change of the average radius of 7.5 mM SDS d/w emulsions for 200, 500, and 700 mM NaCl. For comparison, data corresponding to 0.5 mM SDS d/w for 200 and 300 mM NaCl is also shown.

Figures 9 and 10 show the results of Tr vs h and Tr vs. t corresponding to one set of d/w emulsions. Despite the complexity of some plots of Tr vs. h, the qualitative form of the curves of Tr vs t is the same unless a substantial crystallization occurs. The initial decrease of the transmittance was unexpected, although it has been previously observed that emulsions with $[NaCl] \geq 700$ mM tend to become opaque instantaneously as soon the electrolyte is added. Curves of Tr vs t suggest that this is a general phenomenon. The turbidity of solutions and emulsions augments after the addition of salt until it reaches a maximum, and then decreases. In the case of emulsions an additional decrease in the intensity of the transmitted light could come from structural changes of the surfactant at the interface of the drops [Diaz, 2019].

Since the initial decrease of Tr also occurs at low salinities (below 400 mM NaCl) where substantial aggregation of drops does not occur (Fig. 8) and it is also found in the absence of drops (sw systems), we ascribe it mainly to micelle growth (surfactant aggregation). It is well known that micellar solutions scatter laser light, while diluter surfactant solutions do not (Figure 11). Hence, the transmittance of the solutions should be lower above the cmc. Because the amount of micelles is lower at lower salinity, the number of sampled points (Tr values) corresponding to the slope 1, increases as the ionic strength rises.

Table 2 shows the average slopes of Tr vs t for the sections of the curve where it decreases prior to the mini-

Turbidimetric evaluation of the solubilization rate: Dissolution of dodecane nanodrops in 7.5 mM sodium dodecylsulfate solutions at selected sodium chloride concentrations

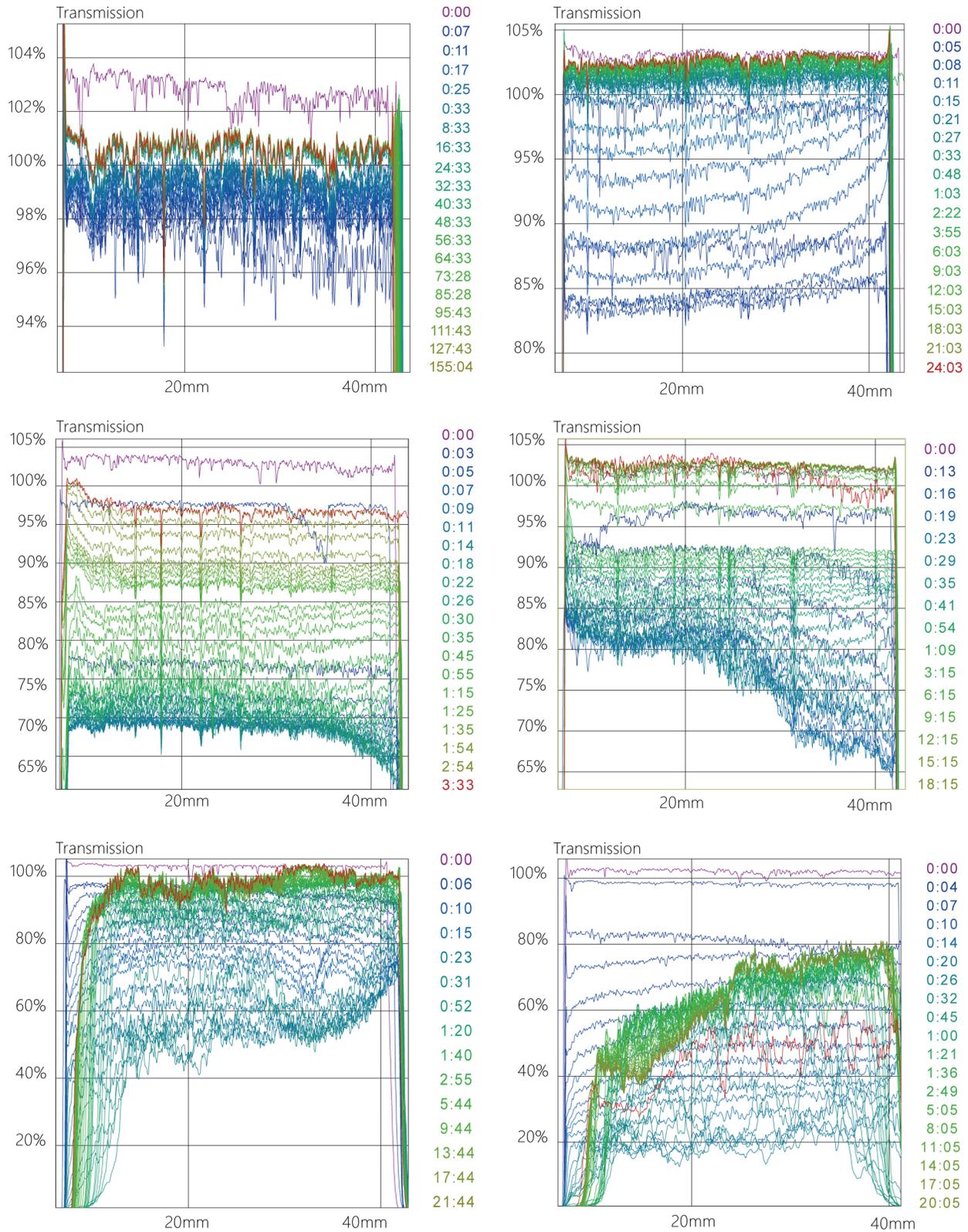

Fig. 9: Evolution of Tr vs. h for d/w emulsions. Illustrations correspond (from top to bottom and left to right) to 100, 300, 500, 700, 900, and 1000 mM NaCl.

José Daniel Rodríguez, Maurice Espinoza, Kareem Rahn-Chique, German Urbina-Villalba

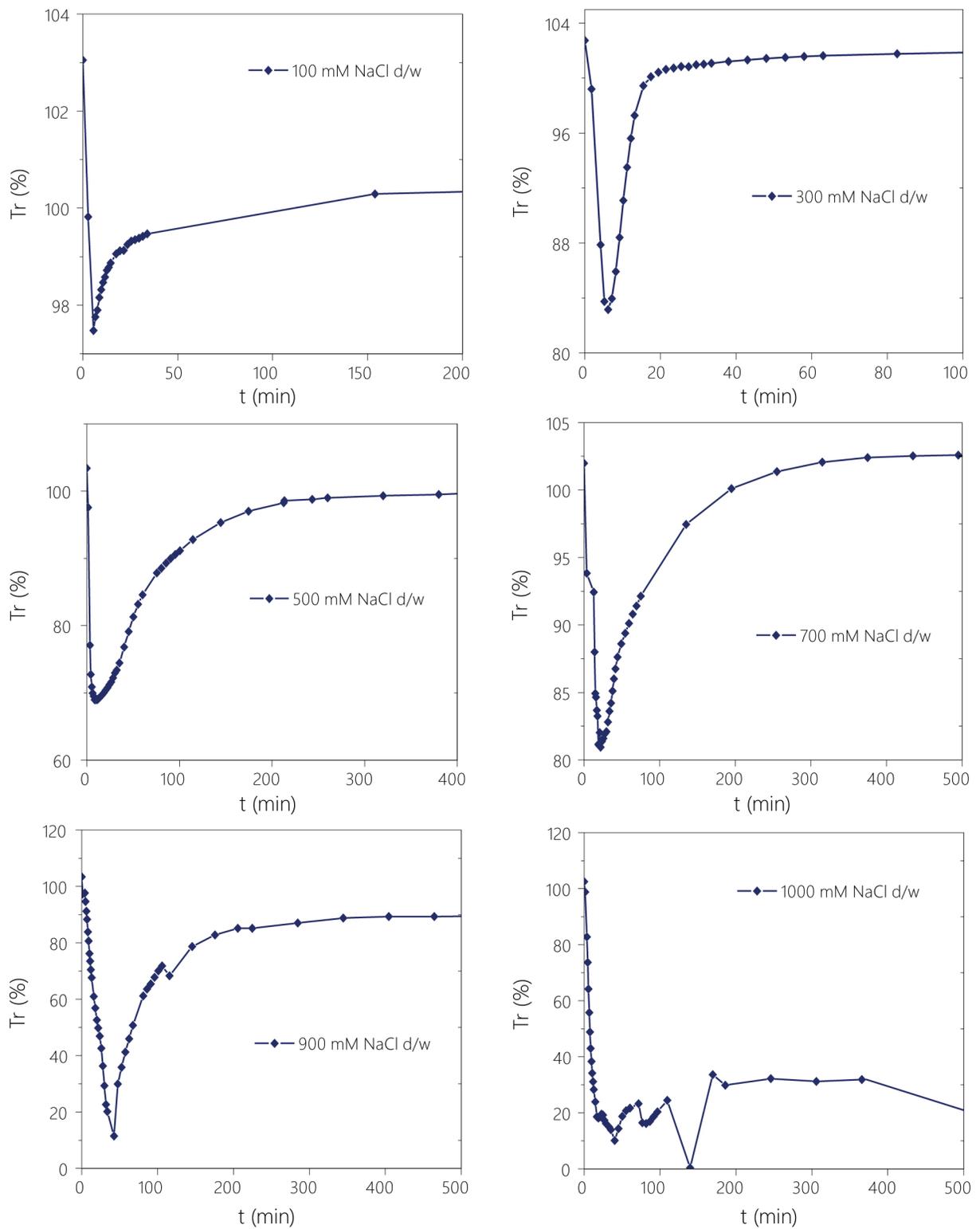

Fig. 10: Plots of Tr vs. t for d/w emulsions corresponding to Fig. 9.

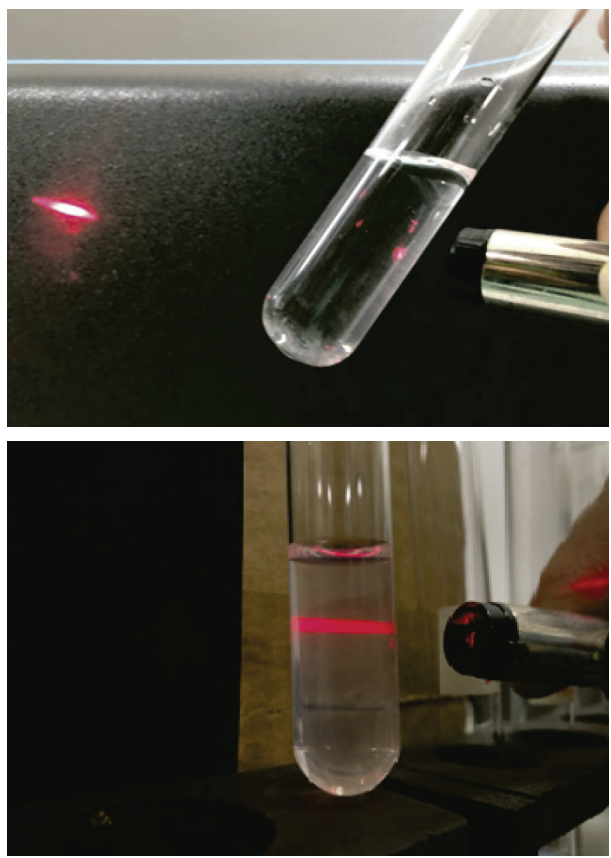

Fig. 11: Laser scattering by sub-micellar and micellar solutions. Light (630 – 670 nm in wavelength, 5 mW) is scattered by micellar structures (around 2 nm) above the cmc. Hence, a bright ray of light is observed crossing the solution (bottom figure). This scattering is not observed below the cmc, since SDS monomers are too small.

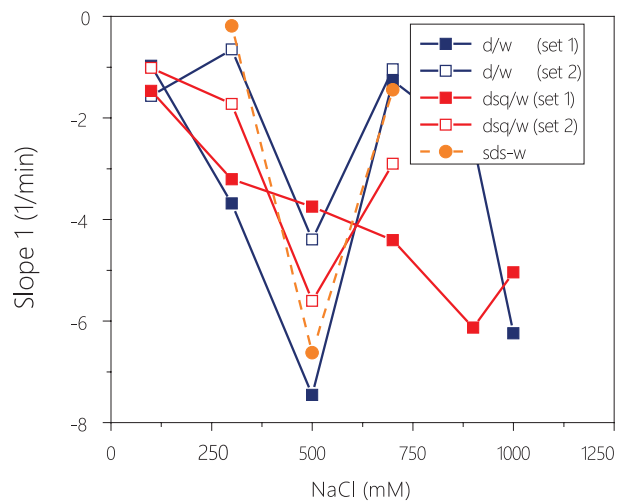

Fig. 12: The minimum rate of micelle formation appears to correspond to 500 mM NaCl.

Table 2: Average slopes of Tr vs. t.

| System | NaCl (mM) | $t_{max}$ (min) | Slope 1 (%Tr/min) | Slope 2 (%Tr/min) | Slope 3 (%Tr/min) |
|---|---|---|---|---|---|
| Dodecane | 100 | 1474 | -1.2733 | 0.1300 | 0.0001 |
| Dodecane | 300 | 1323 | -2.1699 | 1.2599 | 0.0001 |
| Dodecane | 500 | 1340 | -5.9251 | 0.2997 | 0.0004 |
| Dodecane | 700 | 1215 | -1.1535 | 0.4461 | 0.0001 |
| Dodecane | 900 | 1365 | -2.4775 | 0.6186 | -0.0003 |
| Dodecane | 1000 | 1265 | -6.2440 | 0.9336 | -0.0012 |
| Dodecane + Squalene (10%) | 100 | 4151 | -1.2435 | 0.0670 | 0.0025 |
| Dodecane + Squalene (10%) | 300 | 1401 | -2.4665 | 0.0776 | 0.0011 |
| Dodecane + Squalene (10%) | 500 | 1295 | -4.6784 | 0.2642 | 0.0323 |
| Dodecane + Squalene (10%) | 700 | 1335 | -3.6574 | 1.0117 | 0.0244 |
| Dodecane + Squalene (10%) | 900 | 1361 | -6.1320 | 1.8537 | -0.0020 |
| Dodecane + Squalene (10%) | 1000 | 1476 | -5.0412 | 1.8638 | 0.0000 |
| Aqueous Solution of SDS | 300 | 348 | -0.3154 | 0.0801 | 0.0007 |
| Aqueous Solution of SDS | 500 | 1054 | -3.3340 | 1.4780 | 0.0007 |
| Aqueous Solution of SDS | 700 | 1207 | -1.4450 | 0.9804 | 0.5193 |

mum (slope 1), increases just after the minimum (slope 2), and finally, when it asymptotically reaches a stable value of Tr (slope 3). The curvature of Tr vs. t is in general larger for emulsions in comparison to solutions (the absolute value of the slopes is higher). The minimum of the curves is -in most cases- most profound as the salinity increases. There is no monotonous decrease of slope 1 with the amount of salt that could justify the occurrence of flocculation during this initial period. Furthermore: the fact that the slope of the emulsions is minimum for [NaCl] = 500 mM (Fig. 12) as it happens with the surfactant solutions reinforce the idea that the behavior of the system during this period is related to the phase behavior of the surfactant solution. In this regard it should be noticed that surfactant crystallization is only observed for [NaCl] ≥ 500 mM.

Figure 13 shows the curves of Tr vs. h for one set of dsq/w emulsions. The addition of squalene is known to suppress Ostwald ripening, so it could allow identifying the contribution of this process to the overall change of Tr. Remarkably, surfactant crystallization does not occur in the presence of squalene, not even at 1 M NaCl (!). The curves of Tr vs. h and Tr vs. t are smooth and vary monotonously. More strikingly the change of slope 3 for 500 and 700 mM NaCl is linear with a high regression coefficient (> 0.9) suggesting the occurrence of Ostwald ripening (!) within the "slope 3" region

José Daniel Rodríguez, Maurice Espinoza, Kareem Rahn-Chique, German Urbina-Villalba

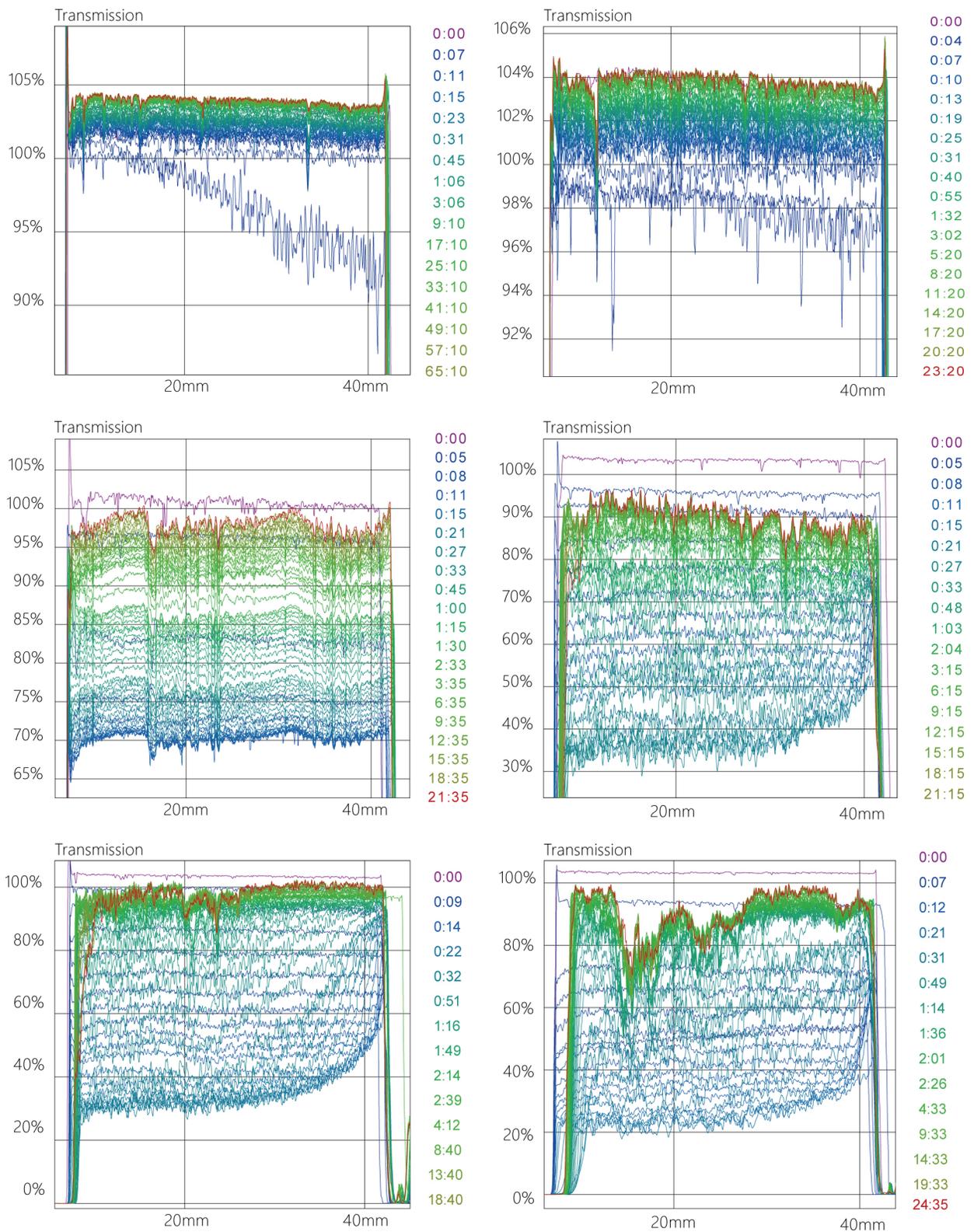

Fig. 13: Evolution of Tr vs. h for dsq/w emulsions. Pictures correspond (from top to bottom and left to right) to 100, 300, 500, 700, 900, and 1000 mM NaCl.

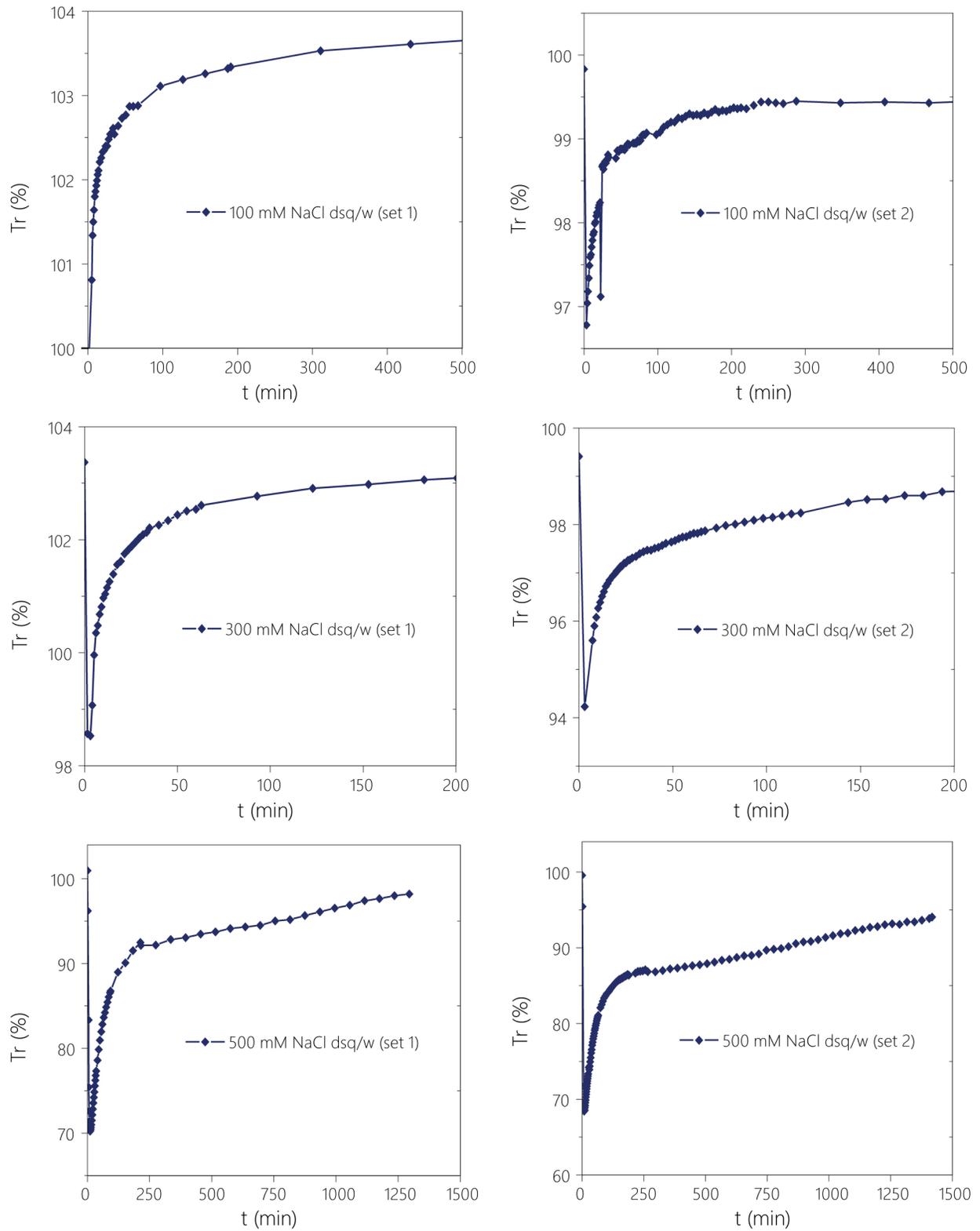

Fig. 14: Plots of Tr vs. t for two sets of dsq/w measurements. First, second, and third row correspond to: 100, 300, 500 mM NaCl, respectively.

José Daniel Rodríguez, Maurice Espinoza, Kareem Rahn-Chique, German Urbina-Villalba

(last time range of our measurements). It is noteworthy that according to our simulations, molecular exchange without complete dissolution of drops causes a decrease of the average radius, which is consistent with an increase of Tr [Urbina-Villalba, 2009].

In order to explore the possible contribution of ripening during the process of solubilization we used Eqs. (11) and (12) to appraise dR/dt, and Eq. (13) to approximate $dR^3/dt$. The results of these calculations are shown in Table 3. According to LSW: $7.3 \times 10^{-28} \leq V_{OR} \leq 2.1 \times 10^{-27}$ m$^3$/s for d/w. Lower values are expected for dsq/w. Except for a few exceptions within the "slope 2" and "slope 3" regions, the calculated values significantly differ from theoretical estimates. The results do not change appreciably for values of m = 1, 3/2, 1/2, or even -3/2. Notice however that the same initial radius of the drop had to be used to parameterize the constants of Eqs. (11) and (12) for sections of different slopes. This is a severe limitation. Moreover, if the process of ripening coincides with solubilization, a rate higher than $V_{OR}$ should be expected. More accurate estimations of $V_{OR}$ are now possible [Rodríguez-López, 2019], but we did not attempt to use more sophisticated theories for this purpose. Equations (11) and (12) only provide an order or magnitude estimation.

Table 4 shows the characteristic times deduced for the calculation of the rates of solubilization ($t_s$), and complete clarification of the sample ($t_c$). As previously discussed, it is likely that the final section of the curve of Tr vs. t is influenced by ripening. If that is the case, the initial rate of increase of Tr just after its minimum value is more likely to reflect the rate of solubilization. Table 5 shows the absolute values of the calculated rates (dR/dt). The value of $t_s$ for 100 mM NaCl is very similar to the one reported by Ariyaprakai and Dungan [Ariyaprakai, 2007; 2008]. Instead, the clarification times are one order of magnitude higher. Clarification times calculated using the criteria of maximum fluctuation of Tr are unreliable (last two columns in Tables 4 and 5) because large fluctuations between two consecutive measurements were found to occur too often, especially in the presence of crystals and within the asymptotic trend of Tr. As a result, several rates correspond to the maximum observation time which is arbitrary. On the other hand, if the fluctuation is calculated with respect to the last value of Tr measured, the values of $t_{0.5}$, $t_{1.0}$ resemble those of $t_s$ and $t_c$.

Table 3: Approximate evolution of the cube average radius according to Eqs. (11), (12) and (13).

| System | NaCl (mM) | Eqs. | m | dR$^3$/dt (m$^3$/s) | dR$^3$/dt (m$^3$/s) | dR$^3$/dt (m$^3$/s) |
|---|---|---|---|---|---|---|
| Dodecane | 100 | 11, 13 | 3 | 4.8 x 10$^{-26}$ | -4.8 x 10$^{-27}$ | -1.9 x 10$^{-30}$ |
| Dodecane | 300 | 11, 13 | 3 | 9.4 x 10$^{-26}$ | -6.6 x 10$^{-26}$ | -3.7 x 10$^{-30}$ |
| Dodecane | 500 | 11, 13 | 3 | 1.9 x 10$^{-25}$ | -1.2 x 10$^{-26}$ | -1.2 x 10$^{-29}$ |
| Dodecane | 700 | 11, 13 | 3 | 3.6 x 10$^{-26}$ | -1.9 x 10$^{-26}$ | -2.9 x 10$^{-30}$ |
| Dodecane | 900 | 11, 13 | 3 | 8.3 x 10$^{-26}$ | -1.8 x 10$^{-25}$ | 1.1 x 10$^{-29}$ |
| Dodecane | 1000 | 11, 13 | 3 | 2.1 x 10$^{-25}$ | -1.7 x 10$^{-25}$ | 4.0 x 10$^{-28}$ |
| Dodecane/Squalene | 100 | 11, 13 | 3 | 3.9 x 10$^{-26}$ | -2.1 x 10$^{-27}$ | -8.1 x 10$^{-29}$ |
| Dodecane/Squalene | 300 | 11, 13 | 3 | 8.4 x 10$^{-26}$ | -2.6 x 10$^{-27}$ | -3.7 x 10$^{-29}$ |
| Dodecane/Squalene | 500 | 11, 13 | 3 | 1.6 x 10$^{-25}$ | -1.2 x 10$^{-26}$ | -1.2 x 10$^{-27}$ |
| Dodecane/Squalene | 700 | 11, 13 | 3 | 1.2 x 10$^{-25}$ | -9.2 x 10$^{-26}$ | -7.5 x 10$^{-28}$ |
| Dodecane/Squalene | 900 | 11, 13 | 3 | 2.0 x 10$^{-25}$ | -1.9 x 10$^{-25}$ | 6.9 x 10$^{-29}$ |
| Dodecane/Squalene | 1000 | 11, 13 | 3 | 1.8 x 10$^{-25}$ | -2.7 x 10$^{-25}$ | 1.3 x 10$^{-30}$ |
| Dodecane | 100 | 12, 13 | 3 | 1.0 x 10$^{-23}$ | -1.6 x 10$^{-25}$ | -1.9 x 10$^{-28}$ |
| Dodecane | 300 | 12, 13 | 3 | 1.1 x 10$^{-23}$ | -3.7 x 10$^{-25}$ | 6.2 x 10$^{-28}$ |
| Dodecane | 500 | 12, 13 | 3 | 4.3 x 10$^{-24}$ | -3.4 x 10$^{-26}$ | -3.9 x 10$^{-28}$ |
| Dodecane | 700 | 12, 13 | 3 | 4.9 x 10$^{-25}$ | -7.4 x 10$^{-26}$ | 3.9 x 10$^{-29}$ |
| Dodecane | 900 | 12, 13 | 3 | 3.0 x 10$^{-24}$ | -2.3 x 10$^{-26}$ | 6.6 x 10$^{-29}$ |
| Dodecane | 1000 | 12, 13 | 3 | 1.7 x 10$^{-23}$ | -3.7 x 10$^{-26}$ | 4.5 x 10$^{-29}$ |
| Dodecane/Squalene | 100 | 12, 13 | 3 | -2.2 x 10$^{-24}$ | 9.4 x 10$^{-26}$ | -1.1 x 10$^{-26}$ |
| Dodecane/Squalene | 300 | 12, 13 | 3 | 4.2 x 10$^{-24}$ | -1.2 x 10$^{-25}$ | -2.8 x 10$^{-27}$ |
| Dodecane/Squalene | 500 | 12, 13 | 3 | 3.5 x 10$^{-24}$ | -2.8 x 10$^{-26}$ | -8.4 x 10$^{-27}$ |
| Dodecane/Squalene | 700 | 12, 13 | 3 | 2.8 x 10$^{-24}$ | -5.6 x 10$^{-26}$ | -6.7 x 10$^{-27}$ |
| Dodecane/Squalene | 900 | 12, 13 | 3 | 2.7 x 10$^{-23}$ | -8.8 x 10$^{-26}$ | 1.7 x 10$^{-27}$ |
| Dodecane/Squalene | 1000 | 12, 13 | 3 | 2.6 x 10$^{-24}$ | -7.8 x 10$^{-26}$ | 6.4 x 10$^{-30}$ |

## CONCLUSION

The fact that the values of $t_s$ are lower for dsq/w than d/w suggest a significant contribution of ripening during the process of solubilization. Within the limitations imposed by our lack of knowledge of the temperature of the sample tube, the values of $t_s$ corresponding to dsq/w are our best estimation for the rate of micellar solubilization of d/w emulsions. According to the present results, $S_R$ does not change drastically as a function of the ionic strength as was formerly expected, and its value is approximately half of the magnitude previously reported (~ 2.3 x 10$^{-11}$ m/s). In any event, it decreases with the ionic strength up to (1.2 – 1.6) x 10$^{-11}$ m/s.

## ACKNOWLEDGEMENT

The assistance of Dr. Gieberth Rodríguez during the Covid pandemic is gratefully acknowledged.

Table 4: Average times of solubilization and clarification according to different criteria

| System | NaCl (mM) | $t_s$ (s) | $t_c$ (s) | $t_{0.5}$ (s) * | $t_{1.0}$ (s) * | $t_{0.5}$ (s) ** | $t_{1.0}$ (s) ** |
|---|---|---|---|---|---|---|---|
| Dodecane | 100 | 22.93 | 353.92 | 153.92 | 153.92 | 153.92 | 5.98 |
| Dodecane | 300 | 14.17 | 323.33 | 175.65 | 43.02 | 17.47 | 15.47 |
| Dodecane | 500 | 118.19 | 680.00 | 500.00 | 440.00 | 344.22 | 284.22 |
| Dodecane | 700 | 67.36 | 435.10 | 375.10 | 315.00 | 315.10 | 255.10 |
| Dodecane | 900 | 94.63 | 444.55 | 764.57 | 724.57 | 718.24 | 558.24 |
| Dodecane | 1000 | 207.13 | 420.44 | 1265.90 | 1265.90 | 1045.90 | 905.90 |
| Dodecane (set 2) | 100 | 22.53 | 128.60 | 27.93 | 12.44 | 3.40 | 3.06 |
| Dodecane (set 2) | 300 | 21.70 | 423.68 | 79.70 | 28.95 | 4.52 | 3.85 |
| Dodecane (set 2) | 500 | 89.10 | 273.27 | 288.41 | 231.57 | 231.57 | 231.57 |
| Dodecane (set 2) | 700 | 107.75 | 162.35 | 1049.05 | 589.03 | 589.03 | 582.37 |
| Dodecane/Squalene | 100 | 40.88 | 439.60 | 430.97 | 86.87 | 6.23 | 6.23 |
| Dodecane/Squalene | 300 | 33.90 | 304.64 | 260.50 | 82.93 | 5.13 | 4.46 |
| Dodecane/Squalene | 500 | 82.88 | 213.72 | 1235.10 | 1075.10 | 1055.11 | 183.02 |
| Dodecane/Squalene | 700 | 41.34 | 355.17 | 1255.20 | 1155.20 | 955.19 | 635.18 |
| Dodecane/Squalene | 900 | 62.18 | 184.79 | 1080.85 | 1080.85 | 1040.88 | 960.85 |
| Dodecane/Squalene | 1000 | 60.14 | 462.37 | 953.85 | 733.85 | 470.88 | 306.57 |
| Dodecane/Squalene (set 2) | 100 | 41.50 | 207.63 | 75.42 | 24.55 | 3.48 | 3.48 |
| Dodecane/Squalene (set 2) | 300 | 53.20 | 428.40 | 398.37 | 170.22 | 7.27 | 7.27 |
| Dodecane/Squalene (set 2) | 500 | 78.80 | 129.30 | 1326.40 | 1236.40 | 752.34 | 44.00 |
| Dodecane/Squalene (set 2) | 700 | 111.23 | 226.53 | 21.15 | 18.38 | 927.18 | 31.80 |

\* with respect to the last measured value; ** between two consecutive mesurements.

Table 5: Average solubilization and clarification rates.

| System | NaCl (mM) | $(dR/dt)_s$ m³/s | $(dR/dt)_c$ m³/s | $(dR/dt)_{t_{0.5}}$ * m³/s | $(dR/dt)_{t_{1.0}}$ * m³/s | $(dR/dt)_{t_{0.5}}$ ** m³/s | $(dR/dt)_{t_{1.0}}$ ** m³/s |
|---|---|---|---|---|---|---|---|
| Dodecane | 100 | 4.4 x 10⁻¹¹ | 2.0 x 10⁻¹² | 4.4 x 10⁻¹¹ | 1.9 x 10⁻¹¹ | 2.5 x 10⁻¹⁰ | 1.9 x 10⁻¹⁰ |
| Dodecane | 300 | 6.0 x 10⁻¹¹ | 7.9 x 10⁻¹³ | 2.9 x 10⁻¹¹ | 9.1 x 10⁻¹² | 1.6 x 10⁻¹⁰ | 1.5 x 10⁻¹⁰ |
| Dodecane | 500 | 9.3 x 10⁻¹² | 4.2 x 10⁻¹² | 3.2 x 10⁻¹² | 2.6 x 10⁻¹² | 3.7 x 10⁻¹² | 4.2 x 10⁻¹² |
| Dodecane | 700 | 1.1 x 10⁻¹¹ | 6.8 x 10⁻¹² | 2.3 x 10⁻¹² | 1.6 x 10⁻¹² | 2.7 x 10⁻¹² | 1.9 x 10⁻¹² |
| Dodecane | 900 | 1.0 x 10⁻¹¹ | 3.5 x 10⁻¹² | 1.3 x 10⁻¹² | 1.3 x 10⁻¹² | 1.7 x 10⁻¹² | 1.2 x 10⁻¹² |
| Dodecane | 1000 | 4.7 x 10⁻¹² | 8.2 x 10⁻¹² | 7.6 x 10⁻¹³ | 7.6 x 10⁻¹³ | 1.1 x 10⁻¹² | 8.9 x 10⁻¹³ |
| Dodecane/Squalene | 100 | 2.3 x 10⁻¹¹ | 2.4 x 10⁻¹² | 2.5 x 10⁻¹¹ | 7.7 x 10⁻¹² | 2.1 x 10⁻¹⁰ | 2.1 x 10⁻¹⁰ |
| Dodecane/Squalene | 300 | 2.3 x 10⁻¹¹ | 2.0 x 10⁻¹² | 8.7 x 10⁻¹² | 3.4 x 10⁻¹² | 1.8 x 10⁻¹⁰ | 1.6 x 10⁻¹⁰ |
| Dodecane/Squalene | 500 | 1.2 x 10⁻¹¹ | 8.3 x 10⁻¹² | 8.4 x 10⁻¹³ | 7.4 x 10⁻¹³ | 1.4 x 10⁻¹¹ | 6.9 x 10⁻¹² |
| Dodecane/Squalene | 700 | 1.6 x 10⁻¹¹ | 5.1 x 10⁻¹² | 2.5 x 10⁻¹¹ | 2.3 x 10⁻¹¹ | 1.5 x 10⁻¹¹ | 5.7 x 10⁻¹² |
| Dodecane/Squalene | 900 | 1.6 x 10⁻¹¹ | 5.6 x 10⁻¹² | 8.9 x 10⁻¹³ | 8.2 x 10⁻¹³ | 1.0 x 10⁻¹² | 7.1 x 10⁻¹³ |
| Dodecane/Squalene | 1000 | 1.6 x 10⁻¹¹ | 2.2 x 10⁻¹² | 1.3 x 10⁻¹² | 9.2 x 10⁻¹³ | 3.2 x 10⁻¹² | 9.7 x 10⁻¹³ |

\* with respect to the last measured value; ** between two consecutive mesurements.